\documentclass[a4paper,11pt]{article}
\usepackage{pos}

\title{ALICE 3: a next-generation heavy-ion detector for LHC Run 5 and beyond}

\author*{Nicola Nicassio}
\author{on behalf of the ALICE Collaboration}

\affiliation{Dipartimento Interateneo di Fisica ‘M. Merlin’ and Sezione INFN,\\ Via Orabona 4, Bari, Italy}

\emailAdd{Nicola.Nicassio@ba.infn.it}

\abstract{The ALICE Collaboration proposes a completely new apparatus, ALICE 3, for the LHC Runs 5 and 6. The detector consists of a large pixel-based tracking system covering eight units of pseudorapidity, complemented by multiple systems for particle identification, including silicon time-of-flight layers, a ring-imaging Cherenkov detector, a muon identification system, an electromagnetic calorimeter and a forward conversion tracker. Track pointing resolution of better than 10~$\mu$m for $p_T$ >200 MeV/$c$ can be achieved by placing the vertex detector on a retractable structure inside the beam pipe. ALICE 3 will, on the one hand, enable novel studies of the quark-gluon plasma (QGP) and, on the other hand, open up important physics opportunities in other areas of QCD and beyond. The main new studies in the QGP sector focus on low-$p_T$ heavy-flavour production, including beauty hadrons, multi-charm baryons and charm-charm correlations, as well as on precise multi-differential measurements of dielectron emission to probe the mechanism of chiral-symmetry restoration and the time-evolution of the QGP temperature. Besides QGP studies, ALICE 3 can uniquely contribute to hadronic physics, with femtoscopic studies of the interaction potentials between charm mesons and searches for nuclei with charm, and to fundamental physics, with tests of the Low theorem for ultra-soft photon emission. This paper will cover the detector concept, the status of novel sensor R\&D and the resulting physics performance.
}

\FullConference{The European Physical Society Conference on High Energy Physics (EPS-HEP2023)\\
 21-25 August 2023\\
Hamburg, Germany\\}

\begin{document}

\maketitle

\section{Introduction}

ALICE (A Large Ion Collider Experiment) is a general-purpose experiment at the CERN Large Hadron Collider (LHC) optimized to study the physics of ultrarelativistic heavy-ion collisions. The main goal of the ALICE Collaboration is to study the microscopic dynamics of the strongly interacting matter produced in such collisions and, in particular, the properties of the quark-gluon plasma (QGP), a state of matter where quarks and gluons are deconfined. 
The LHC Runs 1 and 2 heavy-ion campaigns have already led to crucial advances in our understanding of the properties of QCD phase diagram and of the QGP \cite{cit_alice_physics_summary}. 
 With the current ALICE 2 and the future ALICE 2.1 upgrades, LHC Runs 3 and 4 will allow further systematic measurements of fundamental properties for our understanding of the QGP \cite{cit_run3_and4_perspectives}. 
However, despite the rich planned physics program, fundamental questions will remain open and addressing such questions requires substantial improvements in terms of detector performance and rate capabilities, calling for a next-generation heavy-ion experiment, the so-called ALICE 3 upgrade planned for LHC Run 5 and beyond \cite{cit_alice_3_LoI}.

 \begin{figure}[!bt] 
\centering\includegraphics[width=0.98\columnwidth]{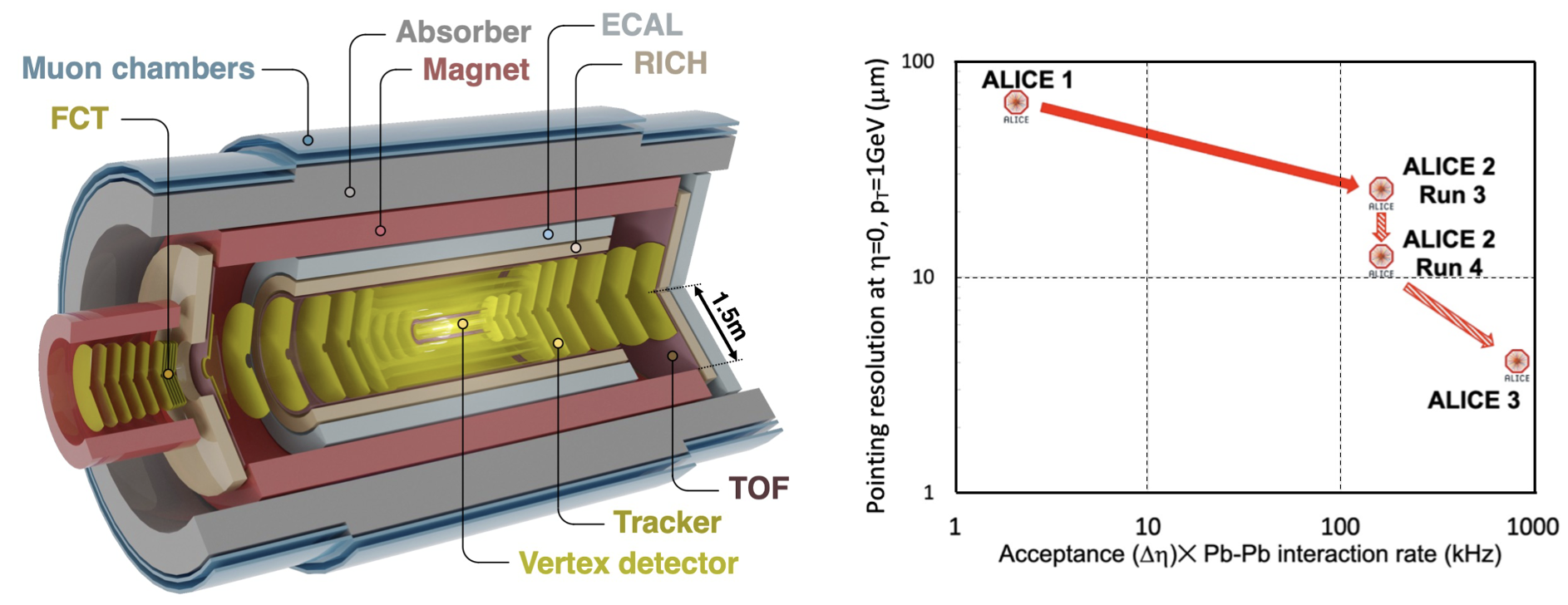}
\caption{Left: Schematic view of ALICE 3 state-pf-the-art detector concept. Right: Improvement in pointing resolution and expected effective statistics in heavy-ion collisions from ALICE 1 to ALICE 3.}
\label{schematic_and_rate_pointing}
\end{figure}
 
\section{ALICE 3 physics goals}

The ALICE 3 primary physics goal is to achieve a deep understanding of the underlying mechanisms affecting the different stages of heavy-ion collisions, from the early-stage medium formation to the heavy-flavour diffusion, thermalization and hadronization in the QGP. 
The main new studies in the QGP sector focus on low transverse momentum ($p_T$) heavy-flavour production, including beauty hadrons, multi-charm baryons and charm-charm correlations, as well as on measurements of net-quantum number fluctuations to constrain the susceptibilities of the QGP and on precise measurements of dilepton emission to probe the time-evolution of the QGP, the early-stage temperature and the mechanism of chiral-symmetry restoration. 
Besides QGP studies, ALICE 3 can uniquely contribute to hadronic physics, with femtoscopic studies of the interaction potentials between charm mesons and searches for nuclei with charm, and to fundamental physics, with precision experimntal tests of Low's theorem for ultra-soft photon emission and searches for axion-like particles (ALP) in ultra-peripheral heavy-ion collisions via light-by-light scattering measurements.

\section{ALICE 3 detector concept and R\&Ds}

All the measurements planned to address ALICE 3 physics goals set strong requirements on vertexing, tracking, particle identification and rate capabilities. The combination of these requirements led to the state-of-the-art detector concept shown in Fig. \ref{schematic_and_rate_pointing}. 
The key requirements are a retractable vertex detector with an unprecedented pointing resolution, a compact and lightweight all-silicon tracker combined with a superconducting magnet system, extensive identifications of $\gamma$, e$^{\pm}$, $\mu^{\pm}$, $\pi^{\pm}$, K$^{\pm}$ and p$^{\pm}$ with different dedicated subsystems, large pseudorapidity acceptance ($|\eta|<4$) and continuous read-out combined with online processing.
The importance of the ALICE 3 upgrade mostly relies on the combined progress in effective statistics (luminosity $\times$ acceptance)  and pointing resolution.  
Fig. \ref{schematic_and_rate_pointing} shows the overall expected improvement by orders of magnitude in the detector capabilities from ALICE 1 (Runs 1 and 2) to ALICE 2 (Run 3), ALICE 2.1 (Run 4) and ALICE 3 (Runs 5 and 6).
In the following, the main features of the different ALICE 3 subsystems are outlined and some of the corresponding ongoing R\&D activities are presented.

\begin{figure}[!tb] 
\centering\includegraphics[width=0.98\columnwidth]{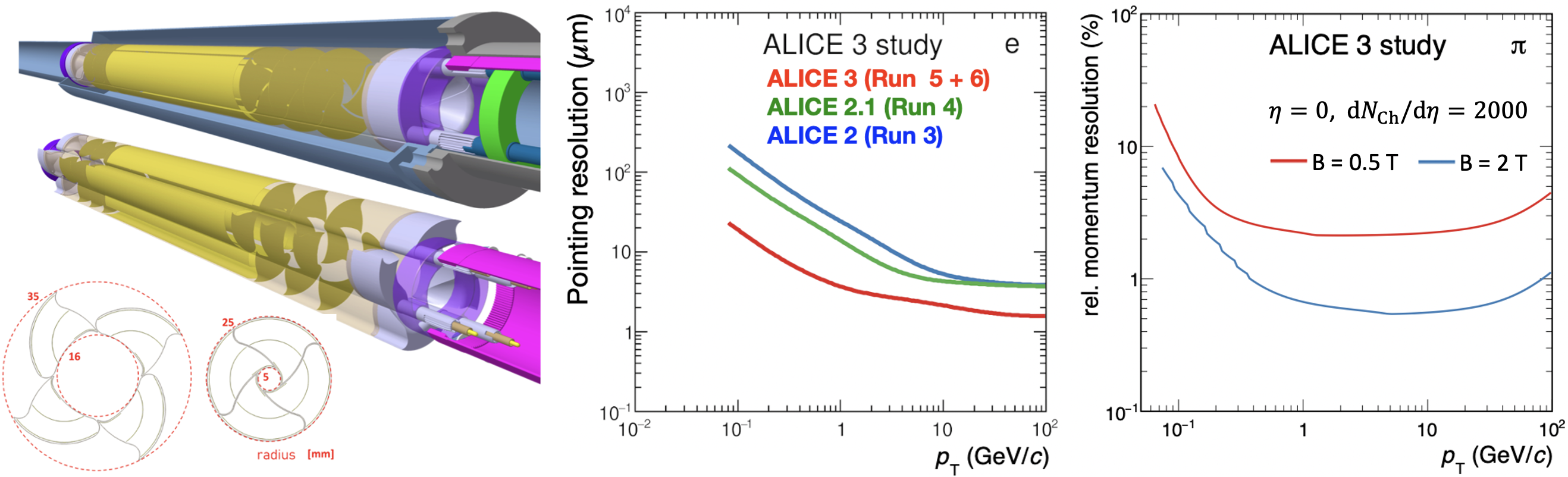}
\caption{Left: Sketch of the ALICE 3 vertex detector for stable beam (top) and at beam injection (bottom). Center: Expected pointing resolution for electrons as a function of $p_{T}$. Right: Expected relative $p_T$ resolution for pions as a function of $p_T$ assuming B = 0.5T (red) and B = 2T (blue). }
\label{vertexing_and_traking}
\end{figure}

\subsection{Vertexing and tracking}

The heart of ALICE 3 is a silicon vertex detector designed to provide a pointing resolution $\sigma_{DCA}$ better than 10 ${\mu}$m for $p_{T}$ larger than 200 MeV. 
The required $\sigma_{DCA}$ can be only achieved by using ultra-thin silicon sensors, featuring an unprecedented low material budget of 0.1\% of a radiation length, and with the first tracking layer placed at a radius of 5 mm from the beam axis at top energy. However, a wider aperture of $\approx$15 mm is required at injection energy, demanding for a retractable detector design.
The current baseline consists of wafer-sized, bent Monolithic Active Pixel Sensors (MAPS) with 10 $\mu$m pixel pitch arranged in 3 barrel layers and 3 forward disks at each end-cap installed in a secondary vacuum inside the beampipe and mounted such that they can be retracted during LHC injection and placed close to the interaction point for data taking, as shown in Fig. \ref{vertexing_and_traking}. The resulting pointing resolution for electrons as a function of $p_T$ is also shown. 
The main R\&D challenges concern mechanical supports, cooling and radiation tolerance of the sensors.

The vertex detector is  complemented by an outer tracker consisting of 8 cylindrical layers and 9 forward disks at each end-cap equipped with  equipped with MAPS having a pixel pitch of 50 $\mu$m, with a material budget thickness of  $\approx$1\% of a radiation length per layer, and installed in a volume of 80 cm radius and $\pm$4 m length around the interaction point. 
The leading requirement for the outer tracker is a relative $p_{T}$ resolution better than 1\% over a wide $p_{T}$ range.  
The momentum is reconstructed from the curvature in a solenoidal magnetic field of B~=~2T provided by a superconducting magnet system featuring an inner radius of 1.5 m. 
Fig. \ref{vertexing_and_traking} shows the expected relative $p_{T}$ resolution for pions at $\eta = 0$ as a function of $p_{T}$.
The implementation of additional forward dipoles to keep an excellent relative $p_{T}$ resolution up to $|\eta| = 4$ is under study.

\begin{figure}[!t] 
\centering\includegraphics[width=0.99\columnwidth]{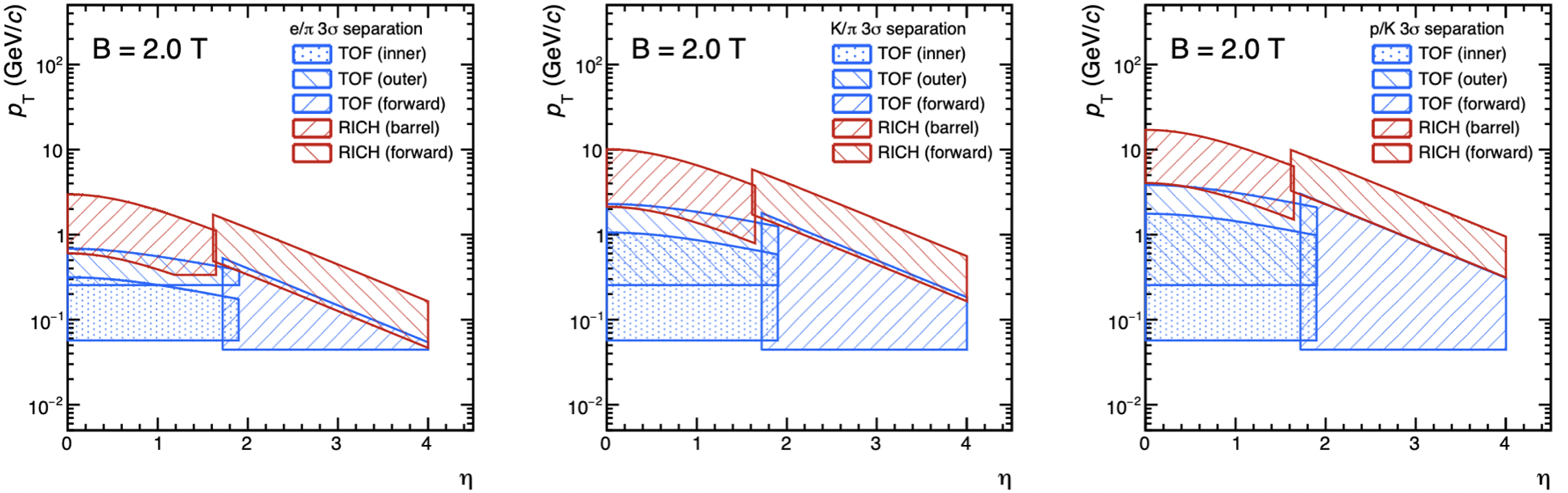} 
\caption{Expected $\eta-p_{T}$ regions in which particles can be separated by at least $3\sigma$ with the ALICE 3 TOF and RICH systems. Electron/pion, pion/kaon and kaon/proton separation plots are shown from left to right.}
\label{pid_coverage}
\end{figure}

\subsection{Particle identification}

The tracker is complemented by dedicated particle identification (PID) systems.  With a time resolution of 20 ps, an inner Time-Of-Flight (TOF) layer (iTOF) at a radius of 20 cm, an outer TOF layer (oTOF) at a radius of 85 cm and forward TOF disks (fTOF) at $\pm4.05$ m distance from the interaction point ensure e/$\pi$, $\pi$/K and K/p separation up to 0.5, 2 and 4 GeV/$c$, respectively, as shown in Fig. \ref{pid_coverage}.
Three sensor technologies have been identified for dedicated R\&Ds: MAPS, Low Gain Avalanche Diodes (LGADs) and Silicon Photomultipliers (SiPMs). 
Recent beam test results prove the feasibility of the 20 ps resolution requirement using LGADs \cite{cit_LGAD_time_resolution} and SiPMs \cite{cit_SiPM_time_resolution}.

The TOF system is complemented by a barrel Ring-Imaging Cherenkov (RICH) 
layer (bRICH) with an inner radius of 90 cm and forward RICH disks (fRICH)
at $\pm4.05$ m distance from the interaction point based on an aerogel radiator  with a refractive index of 1.03 and a SiPM-based photodetector 
providing a Cherenkov angle resolution better than 1 mrad to extend e/$\pi$, $\pi$/K and K/p separation above 2, 10 and 16 GeV/$c$, respectively, as shown in Fig. \ref{pid_coverage}. A first small scale prototype was successfully tested on beam, with an extrapolated angular resolution consistent with the 1 mrad requirement \cite{cit_mio_RICH_test_beam_2022}. The main R\&D challenges include characterization studies on the radiation tolerance of the sensors, cooling and the mechanics of the whole RICH system.

For the identification of photons and high energy electrons, up to hundreds of GeV, an Electromagnetic Calorimeter (ECal)  based on  Pb-scintillator sampling technology and including a high energy and spatial resolution segment based on PbWO$_{4}$ crystals at $\eta=0$ is foreseen behind the bRICH and the fRICH to cover the region $-1.5 < \eta < 4$. The ECal specifications are optimized to enable precision measurements of radiative decays of $\chi_{c}$ and $D^{*0}$ states and photon-jet correlations.

The identification of ultra-soft photons with  $p_{T}$ down to 1 MeV/$c$, is only possible by exploiting the Lorentz boost at large $\eta$ through their conversion to e$^{\pm}$ pairs. A dedicated Forward Conversion Tracker (FCT) consisting of 9 disks equipped with ultra-thin MAPS is foreseen to cover the region $3 < \eta < 5$. A major R\&D challenge is to limit the amount of material in front of the FCT to less than 10\% of a radiation length to suppress the background from bremsstrahlung processes.

For the identification of muons  down to $p_T \approx 1.5$ GeV/$c$, a dedicated Muon Identification system (MID) consisting of a 70 cm thick steel hadron absorber followed by dedicated muon chambers equipped with scintillator bars readout using SiPMs covering up to $\eta = 1.3$ is planned to be installed outside of the  magnet. Alternative options to scintillators, such as multi-wire proportional chambers (MWPCs) and resistive plate chambers (RPCs) are also under study. 

\section{ALICE 3 expected performance}

In this section, a selection of projections for the expected ALICE 3 physics performance are reported highlighting the impact of each subsystem specifications on the ALICE 3 physics program.  


\subsection{Dielectrons}

\begin{figure}[!t] 
\hspace{-10pt}
\centering\includegraphics[width=1.01\columnwidth]{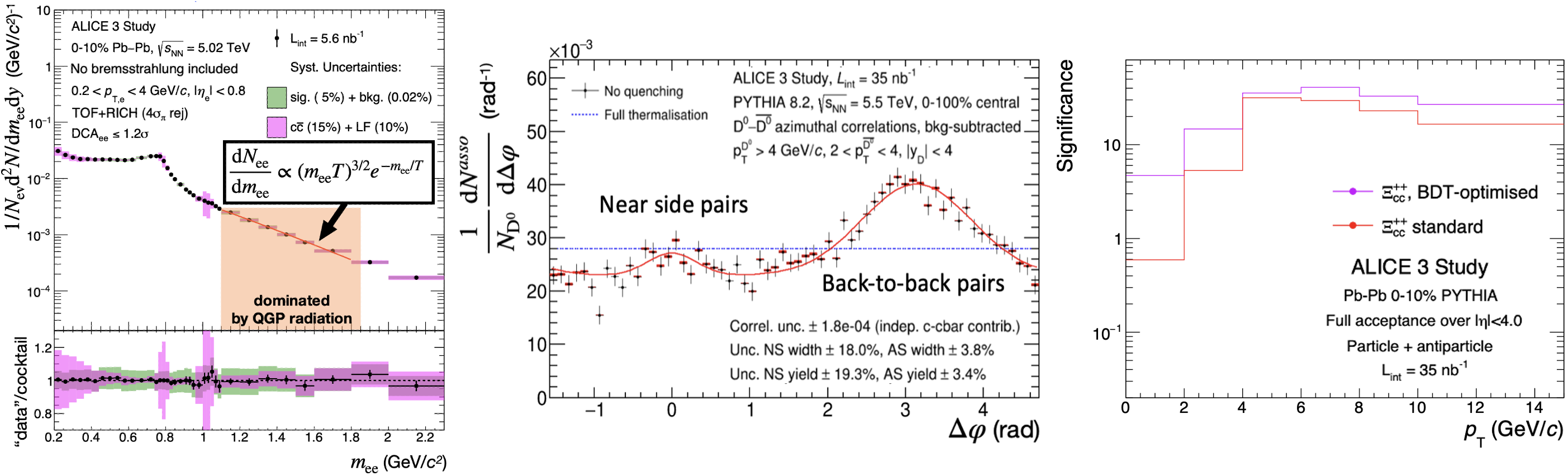} 
\caption{Left: Simulated invariant mass spectrum of thermal dielectrons fitted with an exponential function in the $m_{ee}$ range 1.1-1.8 GeV/$c^{2}$ to extract the early-stage QGP temperature in central (0-10\%) Pb–Pb collisions at $\sqrt{s_{NN}}=5.02$ TeV.
Center: Simulated $D^{0}\bar{D^{0}}$ azimuthal correlation distribution in minimum-bias Pb–Pb collisions with ALICE 3 at $\sqrt{s_{NN}}=5.52$ TeV.
Right: Simulated $\Xi_{cc}^{++}$ significance in central (0-10\%) Pb–Pb collisions at $\sqrt{s_{NN}}=5.52$ TeV as a function of $p_{T}$.} 
\label{physics_projections}
\end{figure}

The hot medium created in heavy-ion collisions emits black-body radiation in form of real and virtual photons. The latter result in dilepton pairs production, representing a powerful probe for QGP temperature and the mechanism of chiral-symmetry restoration studies. Fig. \ref{physics_projections} shows the ALICE 3 projection for the expected thermal dielectron invariant mass ($m_{ee}$) spectrum for central Pb-Pb collisions at a center-of-mass energy per nucleon pair $\sqrt{s_{NN}}=5.02$ TeV after one month of data taking. 
In the region above 1 GeV, the slope of the spectrum provides direct information on the QGP temperature. 
The key ALICE 3 ingredients to achieve the desired statistical and systematic precision are the excellent electron identification down to low $p_{T}$ achieved combining both TOF and RICH information, the small detector material budget in order to reduce the combinatorial background from $\gamma$ conversion or $\pi^{0}$ Dalitz decays and the excellent pointing resolution to efficiently suppress the dominant background from heavy-flavour hadron decays beyond current limits.

\subsection{Heavy-quark correlations}

The angular correlation of heavy-flavour hadron pairs is a powerful probe of QGP scattering, since it is sensitive to the energy loss mechanisms and the degree of heavy-quark thermalization in the QGP, especially at low $p_{T}$. Fig. \ref{physics_projections} shows the ALICE 3 projection for $D^{0}\bar{D^{0}}$ correlation for minimum-bias Pb-Pb collisions at $\sqrt{s_{NN}}=5.52$ TeV after six years of data taking. 
This measurement will be out of reach in LHC Runs 3 and 4, since it requires very high $D^{0}$ purity and reconstruction efficiency, as well as a large $\eta$ coverage, and it will be only possible with ALICE 3.

\subsection{Multi-charm baryons} 

Multi-charm baryon yields are powerful probes of the hadron formation mechanisms since they are expected to be produced by combination of uncorrelated charm quarks produced in heavy-ion collisions. 
The ALICE 3 specifications will enable systematic measurements of multi-charm states currencly inaccessible, such as $\Xi_{cc}^{++}\rightarrow \Xi_{c}^{+}\pi^{+}$ with the subsequent decay $\Xi_{c}^{+}\rightarrow \Xi^{-}\pi^{+}\pi^{+}$.
With the first tracking layer at a distance of 5 mm from the beam axis, strange baryons like $\Xi^{-}$ can be directly tracked before they decay, thus improving the capability to identify weak decays from primary and secondary sources.
The resulting expected significance as a function of $p_{T}$ is shown in Fig. \ref{physics_projections}.

\section{Conclusions}

ALICE 3 is a next-generation detector planned to replace the present ALICE apparatus for LHC Run 5 and beyond. Its main goal is to unravel the microscopic dynamics of the QGP produced in heavy-ion collisions  by exploiting the full potential of the LHC as a heavy-ion collider. Apart from QGP physics, ALICE 3 also addresses fundamental open questions in QCD physics and beyond. An innovative detector concept is needed to to ensure high-precision measurements of heavy-flavour hadrons and electromagnetic probes inaccessible in Runs 3 and 4, demanding for several novel R\&Ds with broad impact also on future high-energy and nuclear physics experiments.

\end{document}